\documentclass{article}
\usepackage[utf8]{inputenc}
\input{epsf}

\usepackage{amsmath}
\usepackage{amssymb}
\input{epsf}
\usepackage{natbib}
\usepackage[margin=1in]{geometry}
\usepackage[raggedright]{titlesec}
\usepackage[hang,flushmargin]{footmisc} 
\usepackage{graphicx}
\usepackage{framed}
\usepackage{color}
\usepackage{yfonts}
\usepackage[dvipsnames]{xcolor}
\usepackage{pifont}
\usepackage{amsmath,amssymb}
\usepackage{multirow}
\usepackage{amsfonts}
\usepackage{subfigure}
\usepackage{caption}
\usepackage{threeparttable}
\usepackage{esvect}
\usepackage{pgf}
\usepackage{tikz}
\usepackage{tikz-cd}
\usepackage{fancyvrb}
\usetikzlibrary{positioning}
\usepackage{bbm,bm}
\usepackage{colortbl}
\usepackage{pdflscape}
\usepackage{booktabs}
\usepackage{mathdots}
\usepackage{yhmath}
\usepackage{cancel}
\usepackage{siunitx}
\usepackage{array}
\usepackage{multirow}
\usepackage{gensymb}
\usepackage{tabularx}
\usepackage{booktabs}
\usetikzlibrary{fadings}
\usetikzlibrary{patterns}
\usetikzlibrary{shadows.blur}
\usetikzlibrary{shapes}
\usepackage{placeins}

\usepackage{hyperref}
\hypersetup{
    colorlinks=true,
    linkcolor=NavyBlue,
    filecolor=magenta,      
    urlcolor=cyan,
    citecolor=black
}
 
\usepackage{rotating}
\usepackage{ntheorem}
\makeatletter
\renewtheoremstyle{plain} 
{\item{\theorem@headerfont ##1\ ##2\theorem@separator}~}
{\item{\theorem@headerfont ##1\ ##2\ (##3)\theorem@separator}~}
\makeatother

\theoremheaderfont{\normalfont\bfseries}

\newtheorem{theorem}{Theorem}[section]

\newtheorem{proposition}{Proposition}[section]

\newenvironment{proof}{\paragraph{Proof:}}{\hfill$\square$}

\newcommand{\E}{\mathbb{E}}

\newcommand{\var}{\text{var}}
\newcommand{\cov}{\text{cov}}
\newcommand{\cor}{\text{cor}}

\definecolor{shadecolor}{gray}{0.9}

\tikzset{every picture/.style={line width=0.75pt}} 

\usepackage{enumitem}
\newlist{Step}{enumerate}{2}
\setlist[Step]{label={{Step \arabic*.}}, leftmargin=*}

\usepackage{setspace}

 \makeatletter
\newcommand\circled[1]{%
  \mathpalette\@circled{#1}%
}
\newcommand\@circled[2]{%
  \tikz[baseline=(math.base)] \node[draw,circle,inner sep=2pt] (math) {$\m@th#1#2$};%
}
\makeatother

 \makeatletter
\newcommand\circledblue[1]{%
  \mathpalette\@circledblue{#1}%
}
\newcommand\@circledblue[2]{%
  \tikz[baseline=(math.base)] \node[draw,circle, fill=blue!20, inner sep=2pt] (math) {$\m@th#1#2$};%
 }

\renewenvironment{abstract}
 {\begin{center}\normalsize\textsc{Abstract}%
 \end{center}\begin{quote}\normalsize}
 {\end{quote}}

\title{Sensitivity Analysis for Clustered Observational Studies}
\author{Melody Huang}
\date{}

\begin{document}

\pagestyle{plain}

\newcommand{\blind}{0}

\newcommand{\tit}{\Large Sensitivity Analysis for Clustered Observational Studies with an Application to the Effectiveness of Magnet Nursing Hospitals}

\if0\blind

{\title{\tit\thanks{This research is supported by the Institute of Education Sciences, U.S. Department of Education, through Grant R305D210014. The opinions expressed are those of the authors and do not represent views of the Institute or the U.S. Department of Education. One of the datasets used for this study was purchased with a grant from the Society of American Gastrointestinal and Endoscopic Surgeons. Although the AMA Physician Masterfile data is the source of the raw physician data, the tables and tabulations were prepared by the authors and do not reflect the work of the AMA. The Pennsylvania Health Cost Containment Council (PHC4) is an independent state agency responsible for addressing the problems of escalating health costs, ensuring the quality of health care, and increasing access to health care for all citizens. While PHC4 has provided data for this study, PHC4 specifically disclaims responsibility for any analyses, interpretations or conclusions. Some of the data used to produce this publication was purchased from or provided by the New York State Department of Health (NYSDOH) Statewide Planning and Research Cooperative System (SPARCS). However, the conclusions derived, and views expressed herein are those of the author(s) and do not reflect the conclusions or views of NYSDOH. NYSDOH, its employees, officers, and agents make no representation, warranty or guarantee as to the accuracy, completeness, currency, or suitability of the information provided here.}}

\author{Melody Huang\thanks{Yale University, New Haven, CT, Email: melody.huang@yale.edu}
\and Eli Ben-Michael\thanks{Carnegie Mellon University, Pittsburgh, PA, Email: ebenmichael@cmu.edu}
\and Matthew McHugh\thanks{University of Pennsylvania, Philadelphia, PA, Email: mchughm@nursing.upenn.edu}
\and Luke Keele\thanks{University of Pennsylvania, Philadelphia, PA, Email: luke.keele@gmail.com}
}

\date{\today}

\maketitle
}\fi

\if1\blind
\title{\tit}
\maketitle
\fi

\begin{abstract}

In a clustered observational study, treatment is assigned to groups and all units within the group are exposed to the treatment. Here, we use a clustered observational study (COS) design to estimate the effectiveness of Magnet Nursing certificates for emergency surgery patients. Recent research has introduced specialized weighting estimators for the COS design that balance baseline covariates at the unit and cluster level. These methods allow researchers to adjust for observed confounders, but are sensitive to unobserved confounding. In this paper, we develop new sensitivity analysis methods tailored to weighting estimators for COS designs. We provide several key contributions. First, we introduce a key bias decomposition, tailored to the specific confounding structure that arises in a COS. Second, we develop a sensitivity framework for weighted COS designs that constrain the error in the underlying weights. We introduce both a marginal sensitivity model and a variance-based sensitivity model, and extend both to accommodate multiple estimands. Finally, we propose amplification and benchmarking methods to better interpret the results. Throughout, we illustrate our proposed methods by analyzing the effectiveness of Magnet nursing hospitals.  


\end{abstract}

\begin{center}
\noindent Keywords:
{Sensitivity Analysis, Clustered Observational Study, Clustered Data}
\end{center}

\clearpage
\doublespacing


\section{Introduction}

\subsection{Magnet Nursing Hospitals}

One key part of health services research is focused on how health care delivery systems affect patient outcomes. For example, do hospitals with superior nursing care have better patient outcomes? One strand of this research has considered whether hospitals with a Magnet Nursing certification have better outcomes for patients. The American Nurses Credentialing Center (ANCC) certifies that hospitals have excellent nursing environments through the Magnet Recognition Program. Hospitals complete an arduous application process and then undergo a peer review that takes approximately one year to complete before they are rewarded with a Magnet certification \citep{aiken1994lower}. Magnet nursing certification has become a widely recognized mark of superior nursing and has become a criterion for ranking in the U.S. News and World Report's Best Hospitals \citep{kutney2015changes}. In terms of whether Magnet hospitals have better patient outcomes, the extant literature has shown that hospitals with Magnet Nursing certification tend to have lower mortality and failure-to-rescue rates as well as better value (lower mortality with similar costs) \citep{aiken1994lower,friese2015hospitals,lake2012association,mchugh2013lower,mitchell1997adverse,silber2016comparison}

Past research on Magnet Nursing hospitals has focused on a range of medical conditions, but has yet to study outcomes for emergency general surgery patients. Emergency general surgery refers to medical emergencies where the injury is internal -- such as a burst appendix or emergency resection for diverticulitis. For the purposes of health service research, researchers have designated 51 medical conditions as EGS conditions that are distinct from trauma conditions \citep{gale2014public,shafi2013emergency}. Care for patients with EGS conditions includes both surgery and a variety of non-operative measures including observation, minimally invasive procedures, medication, and supportive care. EGS conditions are common with an estimated 3-4 million patients per year that account for more than 800,000 operations in the United States alone \citep{gale2014public,scott2016use,havens2015excess,ogola2015financial}. A growing literature has focused on the comparative effectiveness of operative versus non-operative care for various EGS conditions \citep{hutchings2022effectiveness,kaufman2022operative,moler2022local,grieve2023clinical}. Here, we study whether hospitals with Magnet certification have better outcomes for patients with EGS conditions.

In our study, we use a dataset study that merges the American Medical Association (AMA) Physician Masterfile with all-payer hospital discharge claims from Florida and Pennsylvania in 2012-2013. The study population includes all patients admitted for emergency or urgent inpatient care for the 51 recorded EGS conditions. The data also include indicators for the 51 different EGS conditions. The data contain a set of demographic measures including indicators for race, sex, age, and insurance type. The data include indicators for severe sepsis or septic shock and pre-existing disability. Next, there are indicators for 31 comorbidities based on Elixhauser indices \citep{elixhauser1998comorbidity}. Finally, there is an indicator for whether the patient had surgery or not. The data also contain a more limited set of hospital level covariates. Hospital level measures include hospital patient volume, the proportion of emergency admissions, the proportion of patients that had surgery, and the proportion of patients with Medicaid. 

Our focus here is on two different outcome measures. The first outcome is an indicator for the presence of an adverse event following treatment for an EGS condition. The measure of adverse events is an indicator that is 1 if a patient either died, had a complication, or had a prolonged length of stay. The second outcome is a measure of failure to rescue (FTR) -- measured as death after a complication --- and is intended to measure a failure or delay in recognizing and responding to a hospitalized patient experiencing complications from surgery \citep{silber1992hospital}. Across these two states, we have 38 hospitals with Magnet status and 368 hospitals that we designate as controls. In the Magnet hospitals, there are 469,579 EGS patients, and in the control hospitals there are 1,990,315 EGS patients. In Magnet hospitals, 32\% of EGS patients had an adverse event, and in the control hospitals, 27\% of patients had an adverse event. For Magnet hospitals, the FTR percentage was 29\% and for non-Magnet hospitals the FTR percentage was 27\%. These differences are statistically significant but do not include adjustment for baseline confounders and may reflect key differences in the type of hospitals that receive Magnet status rather than any effect of Magnet status itself. Next, we outline a specific research design that we use to answer the question of whether Magnet hospitals offer superior care to EGS patients.

\subsection{Clustered Treatment Assignment}

One common research design in education and health services research is the clustered observational study (COS). In a COS design, an intervention is allocated to groups or clusters of units -- e.g., to hospitals or schools -- while outcomes of interest are measured at the unit level--- e.g., patients or students. There are two critical aspects of treatment assignment in a COS design \citep{Keele:2016b,pagedesign2019,ye2022clustered}. The first critical feature of the COS treatment assignment processes is that all or none of the units within a cluster are exposed to the treatment of interest. The second feature is that treatment assignment is non-random. In health services research, COS designs arise when interventions are applied to entire hospitals or physicians. Such is the case with Magnet Nursing. Here, the intervention is applied to all patients that receive care from a Magnet Nursing hospital and withheld from all patients that visit a non-Magnet hospital. In addition, hospitals are clearly not allocated to Magnet status via a random assignment given that the certification program is highly purposeful.

Given non-random assignment of the treatment in a COS, differences in outcomes may reflect pretreatment differences in treated and control groups rather than actual treatment effects \citep{hansen2014clustered}. For example, Magnet hospitals tend to have larger patient volumes and often are academic medical centers. As such, these factors may account for better patient outcomes rather than any differences in nursing care. As such, consistent with any observational study, we must adjust for observed confounders to make Magnet and non-Magnet hospitals comparable. Statistical adjustment can be done via outcome modeling with regression models. Recent work, however, has focused on the development of specialized forms of statistical adjustment for observed confounders including matching \citep{Keele:2015,Keele:2016b}, the parametric g-formula \citep{ye2022clustered}, and balancing weights \citep{benmichael2022cos}. 

However, all these methods share the same key assumption: that all confounders are measured. Critically, this assumption is untestable, but it can be probed via a sensitivity analysis. A sensitivity analysis is designed to \emph{quantify} the degree to which a key identifying assumption must be violated for an original conclusion to be reversed \citep[ch. 4]{Rosenbaum:2002}. \citet{hansen2014clustered} first outlined a sensitivity analysis for COS designs, but it has two key limitations. First, it is only designed for use with matching estimators, but recent work has developed weighting estimators that can substantially improve on matching methods for adjustment in COS designs \citep{benmichael2022cos}. Second, this form of sensitivity analysis uses a set of worst case bounds that are often overly conservative \citep{huang2022variance}. 

\subsection{Main Contributions and Related Work}

In this article, we develop sensitivity analysis methods for COS designs based on weighting estimators. Currently a variety of approaches have been developed to conduct sensitivity analyses for weighting estimators \citep{tan2006distributional,zhao2019sensitivity,huang2022variance,huang2023design}. However, these methods assume that that treatment assignment is at the unit level; as such, they are not directly applicable to COS applications. The methods we develop are tailored to the balancing weight methods introduced in \citet{benmichael2022cos}, which were developed to adjust for observed confounders in contexts where treatment is assigned to clusters. 

Our contribution is three-fold. First, we derive a bias decomposition framework for clustered treatment assignments. In this framework, we formalize how omitted covariates at either the unit or cluster level can bias treatment effect estimates in the COS design. We highlight the fact that the magnitude of the bias differs depending on whether the omitted covariate is measured at the unit or cluster level. Specifically, we show that omitting cluster level covariates will tend to amplify bias. Second, we develop a sensitivity framework for weighted COS designs that constrain the error that arises in the underlying weights. Specifically, we adapt the marginal sensitivity model \citep{tan2006distributional,zhao2019sensitivity} and the variance-based sensitivity model  \citep{huang2022variance,huang2023design} to account for the COS design. While these approaches have been introduced for the unit level treatment assignment setting, the extension for clustered treatment assignments has not yet been explored. Furthermore, by bounding the error that arises in the underlying weights, the sensitivity models have the flexibility to accommodate multiple estimands beyond the ATE and ATT, such as the average treatment effect in the overlap population, i.e., the ATO \citep{li2018balancing}. This extension is critical to our analysis of Magnet nursing hospitals, since we find limited overlap between the treated and control covariates at the hospital level. In addition, this represents the first extension of sensitivity anlaysis methods to the ATO estimand. 

Finally, to account for the additional complexity in the COS design of having omitted confounders at both the individual and cluster level, we introduce an amplification approach that allows researchers to decompose a single sensitivity parameter into two components: one that controls for the confounding effect from omitting an individual-level confounder, and one that controls for the confounding effect from omitting a cluster-level confounder. We introduce benchmarking approaches that, when used in conjunction with the amplification procedure, can allow researchers to leverage observed confounders to reason about the plausibility of omitted confounders. 

Our article proceeds as follows. In Section~\ref{sec:cos}, we provide a formal description of the COS design using potential outcomes notation. We also review extant balancing weight methods. In Section~\ref{sec:prelim}, we conduct our initial analysis of Magnet hospitals. Since a sensitivity analysis follows estimation of treatment effects, we outline the initial evidence for that Magnet nursing improves outcomes for EGS patients. In Section~\ref{sec:bias} we derive a bias decomposition for clustered treatment assignment. In Section~\ref{sec:sens}, we develop two sensitivity analysis models for the COS design, and apply them to the Magnet nursing estimates.  We then evaluate these methods via simulation in Section~\ref{sec:sim}. In Section~\ref{sec:tools}, we develop tools for interpretation and demonstrate their use with the Magnet nursing sensitivity analysis.

\section{The COS Design}
\label{sec:cos}

\subsection{Notation and Estimands}

We consider a setup with $m$ hospitals, with $n_\ell$ patients in hospital $\ell$, and $n = \sum_{\ell = 1}^m n_\ell$ total patients. We denote the treatment status of cluster $\ell$ as $A_\ell \in \{0,1\}$, $n_1 \equiv \sum_{\ell = 1}^m A_\ell n_\ell$ as the total number of treated units, and $n_0 \equiv n - n_1$ as the total number of control units. Here, when $A=1$, a hospital has magnet status, and when $A=0$ it does not. Each collection of treatment status vectors $\bm{a} = (a_1,\ldots,a_m) \in \{0,1\}^m$ is associated with a vector of potential \emph{cluster assignments}, $\bm{J(a)} = (J_1(\bm{a}),\ldots, J_n(\bm{a})) \in \{1,\ldots,m\}^n$, where $J(\bm{a}) \in \{1,\ldots,m\}$ denotes the cluster that unit $i$ would belong to under overall treatment allocation $\bm{a}$. We denote the observed cluster assignments as $\bm{J} = \bm{J}(\bm{A})$. Each unit $i$ has a \emph{potential outcome} $Y(a_{J(\bm{a})}, \bm{J}(\bm{a}))$ corresponding to the outcome that would be observed if its associated cluster has treatment status $a_{J}$ and the overall cluster assignment is $\bm{J}(\bm{A})$. Note that here we have followed Ye et al \citep{ye2022clustered} and assumed that a patient's potential outcome only depends on its own hospital's treatment status and not the treatment status of other hospitals, except through the potential hospital assignments. 

We further assume that potential outcomes are independent across clusters --- i.e., they are independent for unit $i$ and $i'$ if $J(\bm{a}) \neq J_{i'}(\bm{a})$ --- but may be dependent within clusters. We denote the observed outcomes as $Y(A_{J}, \bm{J})$. We also assume that we observe unit-level covariates $X \in \mathcal{X}$ and that the cluster assignments lead to potential cluster-level covariates, $K_\ell(\bm{J}(\bm(a))) \in \mathcal{K}$, which may include summaries of the unit-level covariates and so can depend on the potential cluster assignments. We let $K_\ell = K_\ell(\bm{J}(\bm(A)))$ denote the observed hospital-level covariates. Next, we use $\mathcal{R} = \{\mathcal{K}, \mathcal{X}\}$ to represent the combined set of patient and hospital-level covariates. Taken together, our observed data consists of tuples $(X, J, K_J, A_{J}, Y)$. We also define two conditional expectations of the observed outcomes. We denote $m_w(a, k) \equiv \E[Y \mid A_{J} = a, K_J = k]$ as the expected outcome for unit $i$, conditioned on its cluster having treatment status $a$ and covariates $k$. We use $m_{wx}(a, k, x) \equiv \E[Y \mid A_{J} = a, K_J = k, X = x]$ to denote the expected outcome if we add more information and additionally condition on unit-level covariates $x$. 

Generally, we focus on the average treatment effect for the treated units (ATT) under the observed cluster assignments $\bm{J}$ as the target causal estimand. We write the ATT in a COS design as: 
\begin{equation}
  \label{eq:att}
  \tau \equiv \underbrace{\E\left[Y(1, \bm{J}) \mid A_{J} = 1\right]}_{\mu_1} - \underbrace{\E\left[Y(0, \bm{J}) \mid A_{J} = 1\right]}_{\mu_0}.
\end{equation}
In our COS design, the ATT estimand measures the effect of treatment for patients in Magnet hospitals, keeping the cluster assignments fixed. The first term, $\mu_1$ can be estimated using the sample average outcome for units in the treated clusters: $\hat{\mu}_1 = \frac{1}{n_1}\sum_{i=1^n}A_{J} Y$. The challenge is identifying and estimating $\mu_0$, the mean counterfactual outcome if those clusters had in fact been assigned to control. Below we review identification conditions for $\mu_0$ under the ATT. However, our methodology readily extends to other common estimands such as the average treatment effect (ATE). Specifically, we also ensure that our methods extend to one estimand known as the average treatment effect for the overlap population (ATO). The ATO is an estimand that measures the treatment effect for the marginal population that might or might not receive the treatment of interest. Critically, under the ATO some treated units receive zero weight. As we outline below, we target the ATO in our analysis of Magnet hospitals given limited overlap between treated and control clusters. Next, we review the two primary COS designs, which rely on different identification assumptions.

\subsection{Identification and Estimation in COS Designs}

There are two primary identification strategies in COS designs, and these different identification strategies will have a direct effect on how we develop sensitivity analysis methods. Specifically, \cite{ye2022clustered} developed two different COS designs: the Cluster-Unit Design (CUD) and the Cluster-Only Design (COD); where each design has different identification assumptions for the same causal estimand. Under the CUD, we must account for differential selection which arises if the mix of units within clusters is changed by the fact that certain clusters are treated. Differential selection would be present in our example if sicker patients seek care at Magnet Nursing hospitals due to the treatment status. As such, with the CUD, we need to correct for unit selection into the clusters, which depends on the treatment assignments. Next, we review the differing identification assumptions for each design.

First, we review identification conditions for the COD. Identification of $\mu_0$ under the COD requires the following set of assumptions:
\begin{itemize}
  \item Cluster assignments are not affected by treatment: $J_\ell(\bm{a}) = J_\ell$ for all $\bm{a} \in \{0,1\}^m$.
  \item For every unit $i$, cluster $\ell$, cluster assignment vector $\bm{j}$, and treatment value $a$, $A_{J} \perp (\bm{J}, X, Y(a, \bm{j})) \mid K_J$.
  \item $P(A_\ell = 1 \mid K_\ell = k) > 0$ for all $k \in \mathcal{K}$.
  \end{itemize}
For the COD, these assumptions form an ignorability assumption based only on cluster-level covariates; there is no need to condition on unit-level covariates. That is, under the COD we assume that there are no unobserved \emph{cluster-level} covariates. In the Magnet hospital application, this would imply that we do not need to condition on \emph{patient-level} covariates. Here, selection into Magnet treatment status is only a function of observed hospital covariates.

Next, identification of $\mu_0$ under the CUD requires the following set of assumptions:
  \begin{itemize}
    \item For two cluster assignment vectors $\bm{j}$ and $\bm{j}'$ such that the treatment assignments are the same, $a_{J} = a_{J}'$, and the cluster-level covariates are unchanged $K_J = K_{j'_i}$ for unit $i$, the potential outcomes are equal $Y(a_{J}, \bm{j}) = Y(a_{j'_i}, \bm{j'})$.
    \item Denote $Y(a, k) = Y(a_{J} = a, K_J = k)$. For every unit $i$, cluster $\ell$, treatment value $a$, and cluster-level covariates value $w$, $A_{J} \perp Y(a, k) \mid K_J, X$.
    \item  $P(A_{J} = 1 \mid K_J = k, X = x) > 0$ for all $k,x \in \mathcal{K,X}$.
    \end{itemize}
For the CUD, the ignorability assumption requires conditioning on both cluster-level covariates and unit-level covariates that drive treatment selection. For the Magnet hospital application, this implies that we need to condition on both patient and hospital level covariates. In fact, the CUD is far more plausible than the COD for Magnet hospitals. That is, it is hard to rule out that some patients visit Magnet hospitals due to the Magnet certification. In addition, the identification assumptions include a version of the stratified interference assumption \citep{hudgens2008toward}. In sum, the key distinction between these two COS designs is the conditioning sets. For the CUD, identification holds if we condition on the combined set of covariates $\mathcal{R}$. For the COD, we only need to condition on the cluster-level covariates $\mathcal{K}$. In a sensitivity analysis, we seek to characterize the bias that results when an investigator omits a covariate from either $\mathcal{R}$ or $\mathcal{K}$. 

Here, we consider estimation of treatment effects for COS designs using weighting estimators. First, we define $w_i$ as a set of estimated weights. Under both the ATE and ATO estimands, the weighting estimator has the form:
\begin{equation}
  \label{eq:w_effect}
  \hat{\tau}(\gamma) = \frac{1}{n_1} \sum_{A_\ell = 1} \sum_{J = \ell} w_i Y - \frac{1}{n_0} \sum_{A_\ell = 0} \sum_{J = \ell} w_i Y.
\end{equation}
For these estimands, we weight both the treated units as well as the control units. Under the ATT, the weighting estimator has the form
\begin{equation}
  \label{eq:w_effect}
  \hat{\tau}(\gamma) = \frac{1}{n_1} \sum_{A_\ell = 1} \sum_{J = \ell} Y - \frac{1}{n_0} \sum_{A_\ell = 0} \sum_{J = \ell} w_i Y.
\end{equation}
Here, we weight only the control units. \citet{ye2022clustered} outlined how to estimate $w_i$ using model-based methods via logistic regression. \citet{benmichael2022cos} developed a balancing weighting estimator for $w_i$ that solves a convex optimization problem to find a set of weights that directly minimize a measure of covariate imbalance subject to an additional constraint or penalty on the complexity of the weights. Both estimation methods, however, generate a set of weights that are used to estimate treatment effects in the same way.

To develop sensitivity analyses, we consider scenarios where the weights $w_i$ are estimated using either $\mathcal{R}$ or $\mathcal{K}$ when the conditioning sets actually have the following form: $\mathcal{R} = \{\mathcal{K}, \mathcal{X}, \mathbf{U}\}$ or $\{\mathcal{K},\mathbf{U}\}$ where $\mathbf{U}$ represents omitted variables or confounders. We compare these misspecified weights to $w_i^*$, a set of ideal weights that would result from conditioning on $\mathcal{R} = \{\mathcal{K}, \mathcal{X}, \mathbf{U}\}$. In this scenario, estimates based on $w_i$ will display some level of bias, while estimates based on $w_i^*$ would be unbiased and consistent. In a sensitivity analysis, investigators use a model to characterize the bias of a weighted estimator based on $w_i$.


\section{The Causal Effect of Magnet Nursing Hospitals}

\label{sec:prelim}
Given that sensitivity analyses are performed after estimating treatment effects, we next estimate the Magnet hospital treatment effect. The first step in our analysis is estimation of the weights. We assume that we are under the CUD design. We adopt the CUD design, since patients and providers may be aware of the Magnet nursing certification and this may shift patient referral patterns. Given this decision, we estimated weights that condition on both patient and hospital level covariates using balancing weights tailored to the COS context \citep{benmichael2022cos}. 

\begin{figure}[htbp]
  \centering
    \includegraphics[scale=0.5]{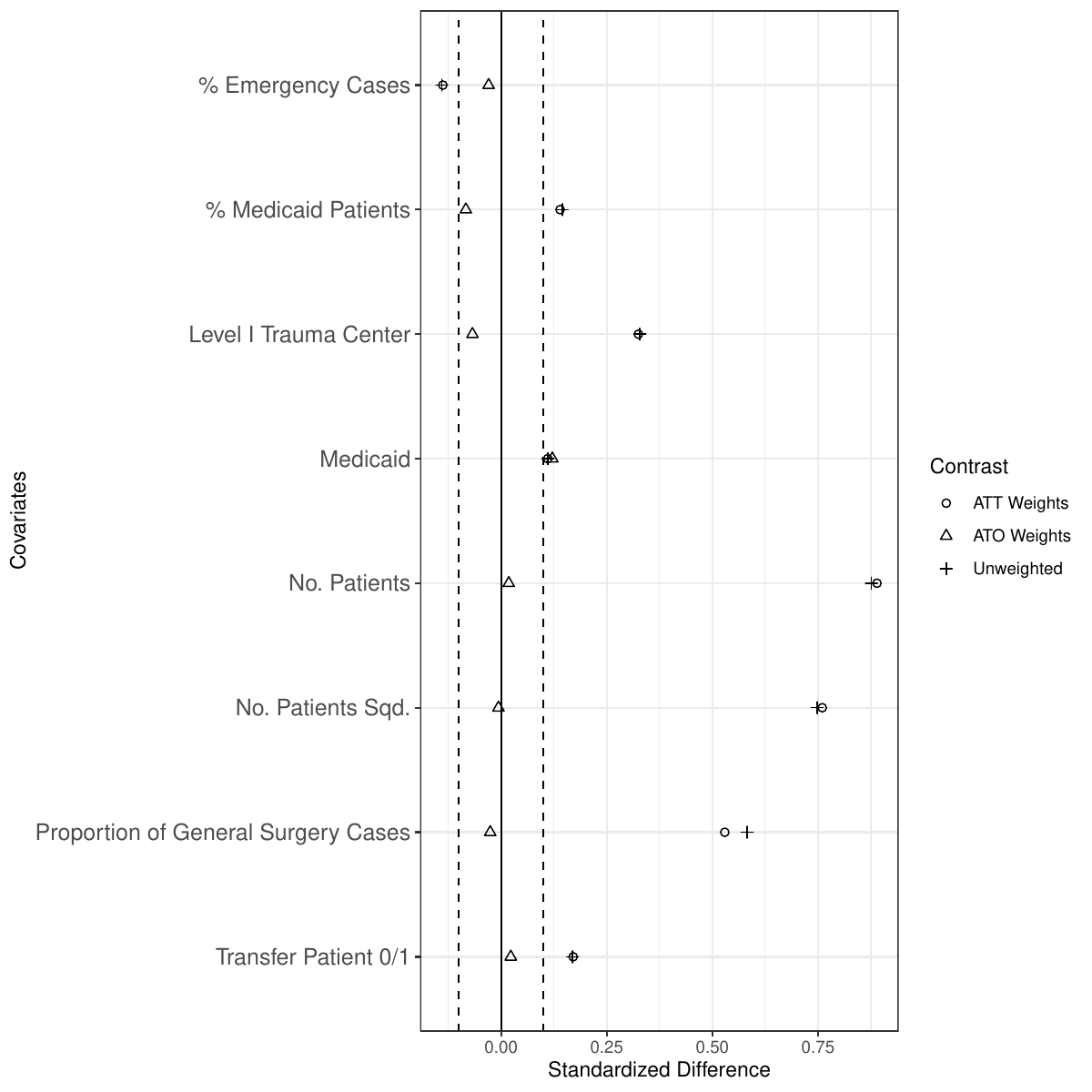}
    \caption{Plot comparing balance measured by standardized difference before and after weighting for the ATT and ATO estimands.}
  \label{fig:bal1}
\end{figure}

First, we estimated weights for the ATT estimand. We selected the hyperparameters using the data-driven procedure outlined in 
 \citet{benmichael2022cos}.  In Figure~\ref{fig:bal1}, we present balance statistics comparing Magnet and non-Magnet hospitals before and after weighting for the subset of covariates with the largest imbalances at baseline. Notably, there are three covariates in particular that are highly imbalanced: the number of patients, the number of patients squared, and the proportion of cases that are admitted for a general surgery procedure. Notably, weights that target the ATT also fail to balance hospitals on these three covariates. This means that Magnet hospitals tend to have much larger patient volumes than non-Magnet hospitals and have a larger share of EGS patients, and there appear to be few non-Magnet hospitals that are comparable on these dimensions. Large imbalances after weighting are often indicative of a lack of overlap between treated and control clusters. 

Given the lack of overlap between the treated and control clusters, we next estimated subset weights that target the average treatment effect ATO COS estimand \citep{benmichael2022cos}. We selected the hyperparameters via the same data-driven process. Figure~\ref{fig:bal1} also contains balance statistics before and after weighting for the ATO estimand. By focusing on the overlap population, we are able to estimate weights that better balance the treated and control groups. For the subset weights, all the standardized differences are now less than 0.10. As such, we are able to render the treated and control groups comparable once we focus on the set of Magnet hospitals that overlap with the control hospitals. Critically, however, it is key to understand that the interpretation of this estimand is quite different. Here, this represents the effect of Magnet nursing hospitals for the subset of Magnet hospitals that overlap with the control hospitals.

Table~\ref{tab:ests} contains the point estimates based on the estimators in \citet{benmichael2022cos}. For the ATT estimand, given the large imbalances, we include additional bias correction via an outcome model. For adverse events, the estimated Magnet Hospital treatment effect is 0.43\% and the confidence intervals include zero. For FTR, the estimated point estimate is -1.9\%, and the confidence interval also includes zero. For the ATO estimand, the estimated Magnet Hospital adverse event treatment effect is -1.7\%. This estimate is statistically significantly different from zero. The FTR treatment effect is -3.3\% and is also statistically significant. Thus, we see that once we adjust for confounders, under the ATO estimand, Magnet Hospitals have lower rates of adverse events compared to non-Magnet hospitals. For the ATT estimand, we observe more limited evidence that Magnet Hospitals have lower rates of FTR. Next, we develop methods to understand whether these conclusions are sensitive to the presence of unobserved confounders.

\begin{table}[htbp]
\begin{center}
\begin{threeparttable}
\caption{The Effect of Magnet Hospitals on Adverse Events and FTR: ATT and ATO Estimands}
\label{tab:ests}
\begin{tabular}{lcc}
\toprule
 & ATT  & ATO   \\
  & Estimand & Estimand  \\
\midrule
Adverse Events  & 0.43 & -1.7 \\
  & $[-1.1, 1.9]$ & $[-3.3, -0.1]$\\
Failure to Rescue  & -1.9 & -3.3 \\
  & $[-3.8, 0.01]$ & $[-5.6, -1.1]$\\  
\bottomrule
\end{tabular}
       \end{threeparttable}
\end{center}
\end{table}

\section{Bias Decomposition for Clustered Treatment Assignments}
\label{sec:bias}

Next, we derive the sources of bias when there are omitted confounders for a clustered treatment assignment. Critically, we show that the amount of bias depends on whether the omitted confounder is at the unit or cluster level. First, we assume that we omit $\{U, V_J\}$, where $U$ is an individual-level covariate, and $V_J$ is a cluster-level covariate. Let $\pi(K_J, X) := \Pr(A_{J} = 1 \mid K_J, X)$ represent the observed population-level propensity score. Furthermore, define $\pi^*(K_J, X, V_J, U) := \Pr(A_{J} = 1 \mid K_J, X,V_J, U)$ as the \textit{oracle} propensity score, since it includes the omitted covariates. We first decompose the bias that arises from omitting $\{V_J, U\}$ for the ATT estimand. We provide the same decomposition for the ATO estimand below, and for the ATE in the supplemental materials.

\subsection{The ATT Estimand}

When the ATT is the target estimand in the CUD, the mis-specified population-level weights have the following form: 
$$
w^{\textsc{ATT}} = \frac{\pi(K_J, X)}{1-\pi(K_J, X)}.
$$
\noindent The correctly specified oracle weights, $w_i^*$ are a function of the oracle propensity scores:
\begin{align} 
w^{\textsc{ATT*}} &= \frac{\pi^*(K_J, X, V_J, U)}{1-\pi^*(K_J, X, V_J, U)} \nonumber \\
&= \underbrace{\frac{\pi(K_J, X)}{1-\pi(K_J, X)}}_{\equiv w_i} \cdot 
\underbrace{\frac{\Pr(V_J \mid K_J, X, A_{J} = 1)}{ \Pr(V_J \mid K_J, X, A_{J} = 0)}}_{\text{Imbalance in Cluster-Level }V_J} \cdot \underbrace{\frac{\Pr(U \mid V_J, K_J, X, A_{J} = 1)}{\Pr(U \mid V_J, K_J, X, A_{J} = 0)}}_{\text{Imbalance in Indiv.-Level } U}
\label{eqn:weight_decomp}
\end{align}

\noindent Close inspection of these weights reveals a number of key insights. First, we note that the error in the weights is driven by the residual imbalance in both $U$ and $V_J$. Second, the imbalance in $U$ (the individual-level bias) depends on how $U$ depends on not only the \textit{observed} covariates $\{K_J,X\}$, but also the omitted cluster-level confounder $V_J$. This result implies that for clustered treatment assignments, we must account for two different types of potential omitted confounders that can result in biased estimates, which is distinct from existing sensitivity approaches for weighting estimators.

We decompose the bias from omitting $\{V_J, U\}$ in the following theorem:
\begin{theorem}[Bias from Omitting Confounders] \label{thm:bias} Let the weights estimated be $w(K_J, X)$. Let $w^*_i \equiv w(K_J, X, V_J, U)$. Now let an intermediary set of weights, in which we omit $U$, but not $V_J$ to be represented as $w(K_J, X, V_J)$. Then, the bias from omitting $\{V_J, U\}$ can be written as: 
\begin{align*} 
\text{Bias}(\hat \tau_{\textsc{ATT}}) =  \sqrt{\frac{1}{1-R^2_V}} \cdot \Bigg[&\underbrace{\cor\left\{w(K_J, X) - w(K_J, X, V_J), Y \mid A_{J} = 0\right\} \sqrt{R^2_V}}_{\text{Bias from cluster-level confounder}}+\\
&~~ \underbrace{\cor\left\{w(K_J, X, V_J) - w^*(K_J, X, V_J, U), Y\mid A_{J} = 0 \right\} \sqrt{\frac{R^2_U}{1-R^2_U}}}_{\text{Bias from unit-level confounder}}~\Bigg]\\
&~~~~~~ \times \underbrace{\sqrt{\var(Y\mid A_{J} = 0) \cdot \var(w(K_J, X\mid A_{J} = 0))}}_{\text{(Estimable) Scaling Factor}},
\end{align*} 
where $R^2_V = \var\left\{w(K_J, X) - w(K_J, X, V_J) \right\}/\var\left\{ w(K_J, X, V_J)\right\}$ represents the residual imbalance in the cluster-level confounder $V_J$, and $R^2_U =  \var\left\{w(K_J, X, V_J) - w^*(K_J, X, V_J, U) \right\}/\var\left\{ w^*(K_J, X, V_J, U)\right\}$ represents the residual imbalance in the individual-level confounder $U$.
\end{theorem} 

\noindent Theorem \ref{thm:bias} decomposes the bias into three parts: (1) the bias from omitting a cluster-level confounder, (2) the bias from omitting a unit-level confounder, and (3) an estimable scaling factor. Next, we provide some interpretation for the key components in this bias decomposition.\\

\noindent \textit{Bias from omitting a cluster-level confounder.} First consider a setting in which we have only omitted a cluster-level confounder and so there is no bias from omitting a unit-level confounder. The resulting bias can then be written as: 
\begin{equation} 
\cor\left\{w(K_J, X) - w(K_J, X, V_J), Y \mid A_{J} = 0\right\} \sqrt{\frac{R^2_V}{1-R^2_V} \cdot \var(Y \mid A_{J} = 0) \cdot \var \left\{ w(K_J, X) \mid A_{J} = 1\right\}}.
\label{eqn:bias_cluster_only}
\end{equation} 
There are parallels between this bias decomposition and the one in standard omitted variable bias settings (e.g., \citealp{huang2022sensitivity, huang2022variance}). In particular, the magnitude of the resulting bias will depend on how much residual imbalance there is in the cluster-level confounder that cannot already be explained by the observed covariates, represented by $R^2_V$, and a correlation term, which captures how related the imbalance is in the cluster-level confounder to the outcome process (i.e., $\cor\{w(K_J, X) - w(K_J, X, V_J), Y \mid A_{J} = 0 \}$). 

The $R^2_V$ value is bounded between $[0,1]$, such that if $R^2_V$ is close to 1, this implies there is greater residual imbalance in the omitted cluster-level confounder. Notably, $R^2_V$ represents the residual imbalance in the omitted confounder that cannot be explained by the observed covariates. For example, in our Magnet Nursing study, one key concern is that we are not fully controlling for hospital quality factors that might be related to both the certification process and patient outcomes. In our analysis, we seek to make hospital comparable in terms of quality, so we controlled for observed covariates such as patient volume and trauma center status. As such, if we believe that much of the imbalance in hospital quality can be explained by patient volume, trauma center status and other hospital level covariates, then the $R^2_V$ value will be relatively small. In contrast, if there is a large portion of the imbalance in hospital quality that cannot be explained by these observed covariates, then there will be a greater degree of residual imbalance, resulting in a relatively high $R^2_V$ value. \\

\noindent \textit{Bias from omitting an individual-level confounder.} Next, we consider the scenario in which there are no omitted cluster-level confounders, but an individual-level confounder has been omitted: 
$$
\cor(w(K_J, X, V_J) - w^*(K_J, X, V_J, U), Y \mid A_{J} = 0) \cdot \sqrt{\frac{R^2_U}{1-R^2_U} \var(Y\mid A_{J} = 0) \var(w(K_J, X)\mid A_{J} = 0)},
$$
Similar to Equation \eqref{eqn:bias_cluster_only}, the bias is driven by an $R^2$ value and a correlation term. $R^2_U$ represents the residual imbalance in an \textit{individual}-level confounder. For example, in our analysis we include covariates to capture the overall patient-mix in each hospital, and in particular try to include controls for the health status of each patient. Now, $R^2_U$ represents the imbalance in patient-mix and health status that cannot be explained by the observed covariates used in the analysis. If this term is large (i.e., $R^2_U$ is close to 1), this implies that even after accounting for the observed covariates, such as comorbidities and demographics, there is still residual imbalance in the unobservable components of the patient mix. In contrast, if $R^2_U \approx 0$, this implies that the observed covariates explain the most of this variation.

It is worth noting that there is an asymmetry in these two bias decompositions. Notably, the bias expression is scaled by $\sqrt{\frac{1}{1-R^2_V}}$. As a result, any imbalance in an omitted cluster-level confounder will \emph{amplify} the bias incurred from omitting an individual-level confounder. Consider a setting in which the omitted cluster-level confounder is highly imbalanced (i.e., $R^2_V$ is close to 1), but is only weakly related to the outcome (i.e., $\cor\{w(K_J, X) - w(K_J, X, V_J), Y \mid A_{J} = 0)$ is close to zero). Then, even if the bias from omitting a cluster-level confounder is relatively small, \textit{any} bias from the unit-level confounder will then be amplified because of the $\sqrt{\frac{1}{1-R^2_V}}$ factor. This same logic applies to observed covariates.  That is, small imbalances in hosptial-level covariates will amplify the bias from imbalances in patient-level covariates. As such, analysts should prioritize balance on the cluster-level covariates when estimating the weights. This point also points to one reason investigators prefer the ATO estimand to the ATT estimand. In our Magnet Nursing analysis, we found it difficult to balance the hospital level covariates when we targeted the ATT. The residual imbalance in hospital level covariates for the ATT estimand may amplify the bias from the imbalance in the patient covariates. However, under the ATO estimand, we are able to achieve far better balance on hospital covariates, which will avoid this kind of bias amplification.\\




\noindent \textit{Scaling Factor.} The final piece of the bias decomposition is $\sqrt{\var(Y \mid A_{J} = 0) \cdot \var(w(K_J, X)\mid A_{J} = 0)}$, which acts as a scaling factor that amplifies the effect of omitted confounders. To provide some intuition, we consider the setting in which $\var(Y \mid A_{J} = 0)$ is large. In this case, there is a high degree of heterogeneity in the control group outcomes that is confounded with the biased treatment assignment process. Similarly, as $\var(w(K_J, X) \mid A_{J} = 0)$ increases, there is greater heterogeneity in the observed covariates, $K_J$ and $X$, for the control group. The presence of such heterogeneity may make it more difficult to balance cluster and individual-level covariates, increasing the potential for bias.

\subsection{The ATO Estimand}

When the ATO is the target estimand in the CUD, the mis-specified population-level weights have the following form: 
$$
w^{\textsc{ATO}} = \begin{cases} 
\pi(K_J, X) & \text{if } A_{J} = 0\\ 
1-\pi(K_J, X) & \text{if } A_{J} = 1\\ 
\end{cases} 
$$
\noindent The correctly specified oracle weights, $w_i^*$ are a function of the oracle propensity scores:
\begin{align} 
w^{\textsc{ATO}*} &= \begin{cases} 
\pi(K_J, X, V_J, U) & \text{if } A_{J} = 0\\ 
1-\pi(K_J, X, V_J, U) & \text{if } A_{J} = 1\\ 
\end{cases} \nonumber \\
&= \begin{cases} 
\displaystyle \pi(K_J, X) \cdot \frac{P(V_J \mid K_J, X, A_{J} = 1)}{P(V_J \mid K_J, X)} \cdot \frac{P(U \mid K_J, X, V_J, A_{J} = 1)}{P(U \mid K_J, X, V_J)} & \text{if } A_{J} = 0\\ 
\displaystyle 1-\pi(K_J, X) \cdot \frac{P(V_J \mid K_J, X, A_{J} = 1)}{P(V_J \mid K_J, X)} \cdot \frac{P(U \mid K_J, X, V_J, A_{J} = 1)}{P(U \mid K_J, X, V_J)} & \text{if } A_{J} = 1\\ 
\end{cases} 
\label{eqn:weight_decomp2}
\end{align}

\noindent Equation \eqref{eqn:weight_decomp2} highlights that we can similarly decompose the oracle ATO weights as a function of the estimated ATO weights $w_i^\textsc{ATO}$ and residual imbalance in the cluster-level omitted confounder, as well as the residual imbalance in the individual-level omitted confounder. 

Next, we require the bias decomposition for the ATO estimand. We include the full bias decomposition in Appendix~\ref{app:bias_ato}.  In that decomposition, we show that the bias for the ATO from omitting $\{V_J, U\}$ decomposes into (1) correlation terms that measure how related the imbalance in the omitted confounders are to the outcomes across treatment and control, (2) how much residual imbalance there is in the cluster-level omitted confounder and the individual-level omitted confounder, and (3) a scaling factor that depends on how much variance there is in the outcomes and estimated weights $w_i^{\textsc{ATO}}$. One advantage of the ATO relative to the ATT (or the ATE) is that the scaling factor in the bias decomposition depends on the variance of the estimated weights. However, since the weights for the ATO are just the propensity scores themselves, they will be lower variance than the weights for the ATT, which are the inverse propensity scores. As a result, the bias from omitting a confounder for the ATO may be smaller than the bias from omitting a similar confounder in the ATT setting. 

\section{Sensitivity Models}
\label{sec:sens}

To assess the sensitivity to omitted confounders, we introduce two different weighted sensitivity models for the COS setting. Following \citet{huang2023design}, we define a \textit{sensitivity model} $\nu$ is defined as a set of hypothetical, ideal weights $w^*$, which is constructed by constraining a local neighborhood around the estimated weights $w$. The size of the neighborhood around $w$ is controlled by a parameter. The larger the parameter, the more we are allowing the ideal weights to differ from the estimated, observed weights.\footnote{Traditionally, the sensitivity models have focused on constraining errors in the underlying propensity scores. In the following paper, we focus on constraining the error in the \textit{weights}.} Thus, as the parameter gets larger, we are allowing for increasing amounts of unobserved confounding. Below, we show that for the COD setting, sensitivity analysis models developed for the unclustered treatment assignment setting can be adapted to COS applications in a straightforward fashion. However, for the CUD, we derive new sensitivity analysis models that allow researchers to consider unit or cluster level confounders, or both. 

Formally, we define a sensitivity model for $\{V_J, U\}$ as $\nu(\Gamma, w(K_J, X)),$ where
\begin{align*} 
\nu(\Gamma, w(K_J, X)) := \Bigg\{ w^*(K_J, X, \{V_J, U\}): f_{\nu}(w(K_J, X), w(K_J, X, \{V_J, U\})) \leq \Gamma \Bigg\}.
\end{align*} 
In words, this sensitivity model defines the set of possible oracle propensity scores $\pi^*(K_J, X, \{V_J, U\})$ that could exist, given a constraint on how different the oracle propensity scores are to the estimated propensity scores. The choice of $f_\nu$ encodes how researchers are choosing to quantify the dissimilarity between the estimated propensity scores and the hypothetical, oracle propensity scores. Different choices of dissimilarity measures correspond to different sensitivity models in the literature. For any given sensitivity model, there is a sensitivity parameter---represented by $\Gamma$---which constraints the total amount dissimilarity the sensitivity model allows for. 

Given a sensitivity model $f_\nu(\Gamma, w(K_J, X))$, researchers can then optimize over the uncertainty set to obtain the partially identified region: 
$$\left[\inf_{\tilde w \in \nu(\Gamma, w)} \tau(\tilde w), \sup_{\tilde w \in \nu(\Gamma, w)} \tau(\tilde w) \right]$$
This partially identified set represents the upper and lower bounds on the treatment effect given the sensitivity model. The minimum value of $\Gamma$ represents the case of no unobserved confounders, and partially identified set will be point identified. Researchers can then increase the values of $\Gamma$ to obtain larger partially identified sets. Investigators are typically interested in finding the value of $\Gamma$ that produces a partially identified region that contains zero. When small values of $\Gamma$ produce a partially identified set that includes zero, this indicates the results sensitive to bias from an unobserved confounder. In the following two subsections, we outline two different sensitivity models for COS designs. The first is the \textit{marginal sensitivity model}, which constrains the worst-case error across the weights, while the second is the \textit{variance sensitivity model}, which constrains the weighted average error.

\subsection{Marginal Sensitivity Model}

We first consider the marginal sensitivity model from \citet{tan2006distributional}. The marginal sensitivity model constrains the worst-case error between the ideal weights and the estimated weights: 
$$
\nu_{\textsc{msm}}(\Lambda, w) = \left \{w^*: \Lambda^{-1} \leq \frac{w(X, K_J)}{w^*(X, K_J, V_J, U)} \leq \Lambda,  \ \ \ \text{for } x \in \mathcal{X}, u \in \mathcal{U}, k_J \in \mathcal{K}_J, v \in \mathcal{V}_J\right\},
$$
where $\Lambda \geq 1$. In other words, the distance measure used to constrain $w$ and $w^*$ is an $L_\infty$ norm.  In the COS design, the marginal sensitivity model from \citet{zhao2019sensitivity}, which assumes that treatment assignment is unclustered, can be directly applied. In the CUD setting, it can be helpful to consider an amplification of the parameter $\Lambda$ (see Section 8 for details). 

For a fixed $\Lambda$ value, researchers can optimize over the set $\nu_{\textsc{msm}}(\Lambda, w)$ to find the maximum and minimum values the estimated treatment effect can take on. In practice, the optimization procedure amounts to solving a linear programming problem \citep{zhao2019sensitivity}. Recent literature has also introduced alternative estimation approaches that allow researchers to construct sharp bounds under the marginal sensitivity model by using a quantile regression \citep[e.g.,][]{dorn2023sharp}.

\subsection{Variance Sensitivity Model}

Bounds based on a marginal sensitivity model tend to be overly conservative in that they result in a partially identified set that is very wide \citep{huang2022variance,huang2023design}. Next, we develop sensitivity methods using a variance sensitivity model. Unlike the marginal sensitivity model, which constrains the worst-case error that arises in the weights, the variance-based sensitivity model constrains a weighted average error. More formally, it will constrain the distributional difference between the hypothetical ideal weights and the estimated weights: 
$$\nu_{vbm} := \left\{ w^*: 1 \leq \frac{\var(w(X, K_J, V_J, U))}{\var(w(X, K_J))}\leq \frac{1}{1-R^2} \right\}$$
An advantage to the variance-based approach is that the constraint on the difference between the variance in the ideal weights and the variance of the estimated weights can be parameterized with respect to an $R^2$ value. We can view the $R^2$ value as how much unexplained variance there is leftover in the ideal weights that cannot be explained by the estimated weights. \citet{huang2022variance} showed that in settings where treatment assignment is unclustered that bounds from a variance sensitivity model were less pessimistic. We will show that using a variance sensitivity model in the COS setting is less pessimistic than the methods based on the marginal sensitivity model.

Another advantage is that we can calculate a closed-form representation of the maximum and minimum bias. For the ATT under the CUD, the maximum bias for a specified $R^2$ value can be written as follows: 
\begin{align*} 
\mathop{\text{max}}\limits_{\tilde w \in \nu_{\text{vbm}}(R^2)}
~\left| \text{Bias}(\hat \tau_{\textsc{ATT}}) \right| 
\leq &\sqrt{1- \cor^2\left\{ w(X, K_J), Y \mid A_J = 0 \right\}} \\
&~~\times \sqrt{\frac{R^2}{1-R^2}\var\left\{ w(X, K_J) \mid A_J = 0\right\} \cdot \var(Y \mid A_{J} = 0)}.
\end{align*} 
This leads to a similar decomposition as the bias decomposition from Theorem \ref{thm:bias}, but bounds the maximum correlation between the imbalance between the estimated and oracle weights and the outcomes. The upper bound is a function of the correlation between the estimated weights $w(X, K_J)$ and the outcomes $Y$. Informally, in settings when \textit{observed} covariates $\{X, K_J\}$ are very related to the outcomes $Y$, then this restricts the amount of residual confounding that can exist from omitted variables $\{U, V_{J}\}$.  This emphasizes a point made in \citet{huang2022sensitivity}, which highlights the importance of weighting on variables that are not only prognostic of the biased selection procedure $A_{J}$, but also the outcome. 

Next, we derive the bound for the maximum bias for a specified $R^2$ value in the ATO setting:
\begin{align*} 
\mathop{\text{max}}\limits_{\tilde w \in \nu_{\text{vbm}}(R^2)}&
~\left| \text{Bias}(\hat \tau_{\textsc{ATO}}) \right| \\
\leq&\overbrace{\sqrt{1- \cor^2\left\{ w(X, K_J), Y \mid A_J = 1 \right\}}
\times \sqrt{\frac{R^2}{1-R^2}\var\left\{ w(X, K_J) \mid A_J = 1\right\} \cdot \var(Y \mid A_{J} = 1)}}^{\text{Maximum bias across treatment units}} \\
&+\underbrace{\sqrt{1- \cor^2\left\{ w(X, K_J), Y \mid A_J = 0 \right\}} 
\times \sqrt{\frac{R^2}{1-R^2}\var\left\{ w(X, K_J) \mid A_J = 0\right\} \cdot \var(Y \mid A_{J} = 0)}}_{\text{Maximum bias across control units}}
\end{align*} 
\noindent The key difference in this bound for the ATO is that we are assuming, for the treatment units, that the error from omitting the confounders is maximally \textit{positively} correlated with the outcomes across treatment. In contrast, for the control units, the error from omitting the confounders will be maximally \textit{negatively} correlated with the outcomes across control. This implies that we are mistakenly overweighting units with larger outcome values in the treatment group, while underweighting units with larger outcome values in the control group. To our knowledge, this the first time such bounds have been extended to the ATO estimand.

\subsection{Inference}

We conclude with a discussion of inferential methods. Both methods provide upper and lower bounds on the true treatment effect given a hypothesized level of confounding. These bounds do not, however, include any statement of statistical uncertainty. As such, inferential methods are needed to provide confidence intervals for the bounds. In settings with unit level treatment assignment, the bootstrap is used for inference \citep{zhao2019sensitivity}. The bootstrap should also provide a method for inference in our setting as well. However, given the clustered nature of treatment assignment, it is necessarily to use the block bootstrap, which resamples the clusters with replacement \citep{Davison:1997}. Below, we evaluate the use of the block bootstrap using simulation.

\subsection{Applying the Sensitivity Analysis for Magnet Nursing}
\label{sec:mag}
Next, we apply these methods to the estimates of the Magnet Nursing treatment effect. In the analysis above, we found that Magnet nursing hospitals reduced FTR under the ATT, and reduced both adverse events and FTR under the ATO. Now, the goal is to understand how sensitive these results are to the possible presence of unobserved confounders. We summarize the results for both the marginal sensitivity model and the variance-based sensitivity models in Table \ref{tab:sens}.

\begin{table}[htbp]
\centering 
\caption{Sensitivity values for Magnet Nursing} \label{tab:sens}
\begin{tabular}{lcccc} \toprule 
& \multicolumn{2}{c}{Adverse Events} & \multicolumn{2}{c}{Failure to Rescue}\\ \midrule 
& MSM & VBM & MSM & VBM \\  \midrule 
ATT Estimand & 1.06 & 0.06  & 1.8 & 0.4 \\ 
ATO Estimand & 1.04 & 0.01 & 1.4 & 0.3\\ \bottomrule 
\end{tabular} 
\end{table} 

We first analyze the results for the marginal sensitivity model. We report the minimum $\Lambda$ value at which the effect bounds include zero. As we outlined above, larger values of $\Lambda$ indicate that it would take larger amounts of unobserved confounding to alter our research conclusions. For the adverse events outcome, the $\Lambda$ values are 1.06 and 1.02 for the ATT and ATO estimands respectively. For example, the estimated treatment effect for adverse events under the ATO is -1.7, and when $\Lambda = 1.02$, the bounds on this point estimate are -3.4 and 1.3. These are small departures from a $\Lambda$ value of 1, the value at which there is no bias from hidden confounders. This indicates that hidden confounding could easily explain this treatment effect. In addition, if we accounted for statistical uncertainty in the bounds, the results would be more conservative. For the FTR outcome, the $\Lambda$ values are 1.8 and 1.4, for the ATT and ATO estimands respectively. Here, the FTR point estimate is -3.3 points under the ATO, and when $\Lambda = 1.8$, the bounds on this point estimate are -6.1 and 0.20. These values of $\Lambda$ are much larger than for the adverse event outcome, indicating that the Magnet Hospital treatment effect on the FTR would require considerably larger amounts of hidden bias from unobserved confounders to explain this result compared to the adverse events outcome, though the amount of hidden bias need not be large in absolute terms.

Next, we report results from the variance sensitivity model, similarly estimating the threshold $R^2$ value at which the effect bounds include zero for both outcomes and estimands. Larger values of $R^2$ correspond to greater distributional differences between the estimated and oracle weights, implying larger amounts of confounding bias. For the adverse events outcome, the $R^2$ values are 0.06 and 0.01 for the ATT and ATO estimands respectively. For the ATO estimand, the bounds on the point estimate are -7.1 and 3.6 at $R^2 = 0.01$. As before, these are quite small values, which indicate that these results could easily be explained by bias from hidden confounders. For the FTR outcome, the $R^2$ values are 0.4 and 0.3 for the ATT and ATO estimands. The bounds on the point estimate for the ATO are -7.2 and 0.50. These values are much larger than those for the adverse event outcome. Overall, these results are consistent with those from the marginal sensitivity model where the results for adverse events are quite sensitive to bias from hidden confounders and the FTR results are more robust.

\section{Simulation Study}
\label{sec:sim}

Next, we conduct two simulation studies to evaluate our proposed methods. We base our simulation study on a design from \citet{benmichael2022cos}, developed to evaluate COS balancing weight methods. First, we outline the data generating process (DGP). The DGP is built from real data designed to evaluate a summer school reading intervention. In the original data, there are 18 treated schools with 1,367 students, and 26 control schools with 2,060 students. The primary outcome in the data is a reading score collected after the summer school reading program. The 5 student-level variables are previous reading and math test scores and indicator variables for race, ethnicity, and sex. The school-level covariates consist of the percentage of students who receive free/reduced price lunch, the percentage who are English language learners, the percentage of students that are proficient in math and in reading, average school level math scores, average school level reading scores, the share of teachers who are novice (e.g., in their first year), the rate of year-to-year staff turnover, and student average daily attendance. 

To simulate from the data, we fit a school-level propensity score model where we regressed the observed treatment indicator on the school level variables. Next, we define the latent probability of treatment as:
\[
Z^* = (\hat{e}(w)/c) + \text{Unif}(-.5,.5)
\]
\noindent where $\hat{e}(w)$ is the estimated propensity score and $c$ controls the level of overlap between treated and control clusters. We then generated treatment assignments via: $Z_j = 1(Z^* > 0.25)$, where $1$ is the indicator function. Next, we generate potential outcomes under control as:
\[
y_0 = \hat{\beta_0} + 2.5 R_{ij} + 2.5 M_{ij} + 1.9 P_{j} + v_1.
\]
\noindent In this model, $R_{ij}$ is student-level reading scores, $M_{ij}$ is student-level math scores, and $P_{j}$ is the percentage of students proficient in math and reading in each school. $\hat{\beta_0}$ is the intercept from an outcome model where we regressed the outcome on the student-level covariates with the basis expanded to include interactions between race and past test scores. Finally, $v_1$ is a draw from a normal distribution that is mean zero with a standard deviation of 12. Next, we generated potential outcomes under treatment as $y_1 = y_0 + \tau$, and we generated simulated outcomes as $\tilde{Y}_{ij} = Z_j y_1 + (1 - Z_j) y_0$. Here, $\tau$ is the true treatment effect estimate in the simulation, and we set it to be 0.3 of a standard deviation of the raw outcome measure. Note that the average ICC across simulations was 0.29 with a standard deviation of 0.04. 

Using this DGP, we conduct a simulation study to understand the performance of our proposed methods and to compare the marginal sensitivity model to the variance sensitivity model. We define $w^*$ as set of weights calculated from the true propensity scores which are based on correctly specified models from the DGP. We consider two settings. First, we consider a setting in which the propensity scores are misspecified with respect to unit (student) level propensity scores. In this case, we omit both $R_{ij}$ and $M_{ij}$ from the estimation of the propensity scores. We then define these weights as $w_u$. Based on $w_u$, we define the oracle sensitivity parameters in this setting as: 
\[
R^2 = 1 - \frac{\var(w_u)}{\var(w^*)}, 
\text{ and }  \Lambda = \max\left\{\frac{w_u}{w^*}, \frac{w^*}{w_u}\right\}, 
\]

Second, we consider a setting when the propensity scores are misspecified from omitting a cluster-level covariate. We define $w_v$ as the set of weights where the school level test scores covariates are omitted from the model. Then, the oracle sensitivity parameters in this setting are defined as: 

\[\ \ 
R^2 = 1 - \frac{\var(w_v)}{\var(w^*)},\ \
\Lambda  = \max\left\{\frac{w_v}{w^*}, \frac{w^*}{w_v}\right\}. \] 

\noindent We use this DGP to conduct two different simulation studies. In the first study, we evaluate whether our proposed sensitivity analysis methods result in valid bounds that contain the true ATT. In the second study, we evaluate the performance of the clustered bootstrap when used for inference. In all the simulations, we focus on the ATT estimand and the corresponding sensitivity analysis for this estimand.

\subsection{Simulation 1}

Our first simulation study is based on the following set of steps: (1) generate data under the DGP, (2) estimate two sets of underspecified COS balancing weights---one with unit variables omitted and one with the cluster variables omitted, (3) compute the sensitivity bounds, and (4) record whether the true treatment effect falls within the sensitivity bounds and the length of the sensitivity bound. In the fourth step, we estimated four different sets of bounds, applying both the marginal- and variance-based bounds for the weights misspecified with respect to student-level covariates and school-level covariates using the oracle sensitivity parameters. In sum, we seek to understand the behavior of the sensitivity analysis when the COS balancing weights are intentionally misspecified, testing whether the sensitivity bounds indeed include the true treatment effect at the oracle values of the sensitivity parameters. We also compare the length of the two sets of bounds to understand the extent to which the marginal model is more conservative.

\begin{figure}
  \centering
    \includegraphics[scale=0.7]{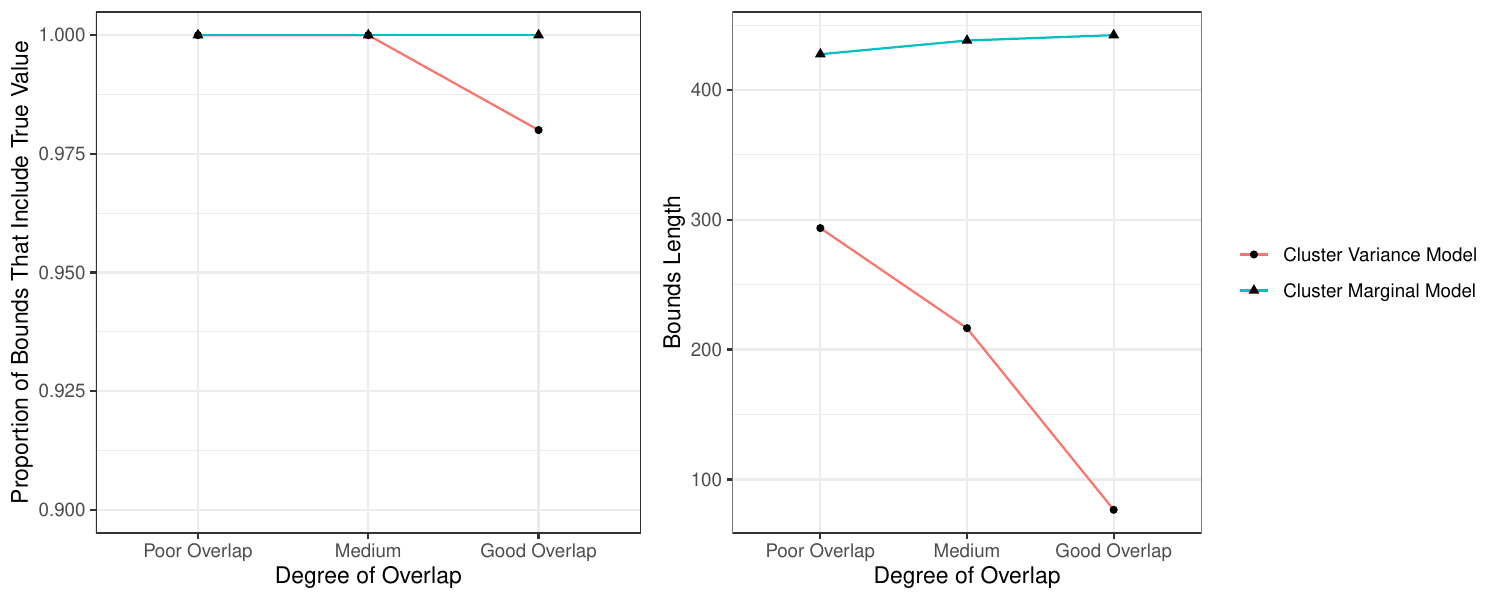}
    \caption{Proportion of times that the sensitivity bounds include true value and length of the sensitivity bounds for three different levels of covariate overlap when cluster-level variables are omitted.}
  \label{fig:sim1}
\end{figure}

In the simulation, we vary the overlap parameter $c$, using values of 1, 5, and 10 which represent poor, moderate and good overlap.
We use 1000 simulations for each value of $c$. We first report the results for when the cluster-level covariates are omitted from the estimated weights in Figure~\ref{fig:sim1}. First, we find that the bounds based on the marginal model include the true treatment effect 100\% of the time across all three levels of overlap. For the variance model, the bounds include the true treatment effect 100\% of the time for two levels of overlap, but the proportion falls slightly below 100\% in one condition. However, there are stark differences in terms of the length of the bounds. First, we observe that the sensitivity bounds based on the marginal model are always wider than the bounds based on a variance model. The length of the bounds for the marginal model are roughly constant across levels of overlap. However, the bounds for the variance model become much shorter as overlap improves. However, even when the bounds from the variance model are at their longest, they are still 25\% shorter than those from the marginal model. The average length of the bounds for the variance model are lowest in the good overlap condition. This most likely explains why the true bounds do not include the true value 100\% of the time in this condition. Overall, the variance model bounds provide good performance but are considerably more informative.

\begin{figure}
  \centering
    \includegraphics[scale=0.7]{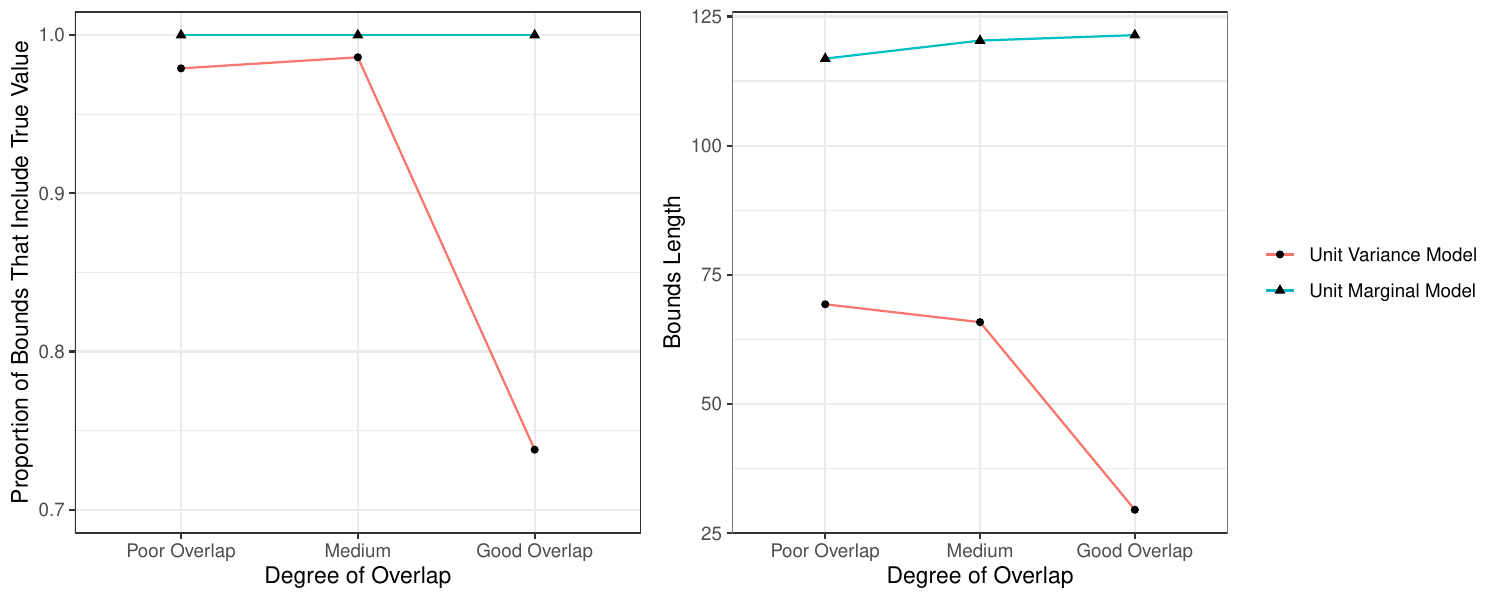}
    \caption{Proportion of times sensitivity bounds include true value and length of sensitivity bounds for three different levels of covariate overlap when unit-level variables are omitted.}
  \label{fig:sim2}
\end{figure}

Next, we report the results for when the unit-level covariates are omitted from the estimated weights in Figure~\ref{fig:sim2}. Again, we find that the marginal model bounds cover the true treatment effect 100\% of the time. However, for the marginal model, the bounds include the true value close to 100\% for two levels of overlap, but it drops below 80\% for one scenario. This result is not too surprising given that the student level variables do not appear in the propensity score model in the DGP. As such, dropping these variables does not change the propensity score much, which implies that the bounds are a function of the variation in the cluster-lwevel model. In terms of the length of the bounds, again those from the marginal model are much wider and do not narrow as the degree of overlap improves. Again, we observe that the variance model produces good performance, but is less conservative.

\subsection{Simulation 2}

Above, we noted that the bootstrap is one way to conduct inference for either type of bounds. In the second simulation, we evaluate the use of the bootstrap as an inferential method for the sensitivity bounds. The basic pattern of this second simulation largely mirrors the first simulation. One key difference is that we do not vary the overlap, only focusing on the case where $c=10$ where we observe some undercoverage of the bounds. For each draw from the DGP, we estimate both the marginal and variance-based bounds at the oracle values for the sensitivity parameter. We then re-sample from the data from this draw from the DGP, and we re-estimate both sets of bounds $B$ times. We then use the percentiles from the bootstrap distributions as confidence intervals for the bounds. Given the clustered nature of the data, we use the block bootstrap which resamples the clusters with replacement but then retains all the units within the resampled clusters without further resampling \citep{Davison:1997}. As an outcome metric, we recorded empirical coverage of the true treatment effect using the bootstrap confidence intervals. In this study, we used 500 simulation replications, and for each of these replications, we resampled 1000 times for the block bootstrap. 

First, we report results for the marginal model. We found in the last simulation that the estimated bounds, under both unit and cluster misspecification, covered the true treatment effect 100\% of the time. The bootstrap confidence intervals also covered the true treatment effect 100\% of the time. Next, we report results for the variance model. When a cluster level covariate is omitted, the block bootstrap confidence intervals cover the true treatment 100\% of the time. When a unit level covariate is omitted, the block bootstrap confidence intervals cover the true treatment effect 98\% of the time. As such, the undercoverage we observed in the prior simulation is remedied by using the block bootstrap for inference.

\section{Sensitivity Analysis Tools}
\label{sec:tools}

Next, we develop two additional tools that can help analysts better understand the results from a sensitivity analysis. These two tools are known as amplification and benchmarking and are widely used in settings with unclustered treatment assignment \citep{huang2023design}. Here, we develop these methods for the COS design. We then apply them to the Magnet Nursing data.

\subsection{Amplification}

Thus far, we have considered the aggregate impact of omitting either a cluster-level or individual-level confounder. In practice, it can be helpful to consider the impact of omitting a cluster-level confounder and an individual-level confounder separately. Next, we outline this type of disaggregation, which is called an amplification. Following our earlier results, we outline amplification for both the marginal and variance-based sensitivity analyses.

First, for the marginal model, following Equation \ref{eqn:weight_decomp}, we can write the multiplicative error between $w$ and $w^*$ as a product of two imbalance terms: the imbalance in the cluster-level $V_J$, and the imbalance in the individual-level $U$. 
\begin{proposition}[Amplification of $\Lambda$] Define $\Lambda_V$ and $\Lambda_U$ as: 
$$\Lambda^{-1}_V \leq \frac{w(X, K_J)}{w(X,K_J, V_J)} \leq \Lambda_V \text{ and } \Lambda^{-1}_U \leq \frac{w(X, K_J, V_J)}{w(X,K_J, V_J, U)} \leq \Lambda_U,$$
for $x \in \mathcal{X}, u \in \mathcal{U}, k_J \in \mathcal{K}_J, v \in \mathcal{V}_J$. Then, we can amplify $\Lambda$ into the product of $\Lambda_V$ and $\Lambda_U$: 
$$\Lambda := \Lambda_V \cdot \Lambda_U.$$
\end{proposition} 
$\Lambda_V$ constrains the maximum imbalance in the cluster-level confounder, while $\Lambda_U$ constrains the maximum imbalance in the individual-level confounder. 

Next, for the variance-based sensitivity model, we can amplify the $R^2$ value into two components that control the imbalance in the omitted cluster-level confounder and the imbalance in the individual-level confounder:

\begin{proposition}[Amplifying $R^2$] \label{prop:amplify_r2} Define $R^2_V$ and $R^2_U$ as follows:
$$\frac{\var(w(X, K_J) \mid A = 0)}{\var(w(X, K_J, V_J) \mid A = 0)} \leq 1-R^2_V, \text{ and } \frac{\var(w(X, K_J, V_J) \mid A = 0)}{\var(w(X, K_J, V_J, U) \mid A = 0)}\leq 1-R^2_U,$$
for $x \in \mathcal{X}, u \in \mathcal{U}, k_J \in \mathcal{K}_J, v \in \mathcal{V}_J$. Then, we can amplify $R^2$ as 
$$R^2 = 1- (1-R^2_V) \cdot (1-R^2_U).$$
\end{proposition}

The amplification of the sensitivity parameter into two components allow researchers to explore the effects of cluster-level and individual-level confounders separately. This is helpful in settings where researchers have a strong substantive priors about the types of confounders they are most worried about. For example, if they are more concerned with potentially omitting a cluster-level confounder (as opposed to an individual-level confounder), the amplification allows them to disentangle the two from one another. Furthermore, amplification allows researchers to evaluate how the resulting bias changes as the cluster-level confounder (or the individual-level confounder) varies in strength (relative to the other). Finally, amplification is useful in concert with the benchmarking procedure we outline next (see Section \ref{sec:bench}), which we can use to calibrate potential sensitivity parameter values. 

\subsection{Benchmarking}
\label{sec:bench}

Researchers can use a benchmarking analysis to calibrate sensitivity parameter values. Specifically, benchmarking allows researchers to compare sensitivity parameters associated with an omitted confounder to the confounding strength of observed covariates. Such an analysis proceeds by sequentially omitting cluster-level covariates and re-estimating the weights. A similar exercise can be done for the individual-level covariates. Researchers can additionally omit subsets of observed covariates (i.e., an individual-level covariate, and a cluster-level covariate) to estimate the bias that would occur from omitting different types of confounders. In each case, one can compare these under-specified weights to the fully specified weights to understand the effect of each type of under-specification.

To formalize benchmarking, we first define what we mean for an omitted confounder to have equivalent confounding strength to an observed covariate. For the variance based sensitivity model, this means that the distributional difference that arises in the estimated and oracle weights is equivalent to the distributional difference that arises in the weights with and without the benchmarked covariates. More formally, let $\{X^{(b)}, K_J^{(b)} \} \in\{ X, K_J\}$ represent the set of individual-level and cluster-level covariates we are trying to benchmark. Then, defining $\tilde X = X \backslash X^{(b)}$ and $\tilde K_J \backslash K_J^{(b)}$, a set of omitted confounders $\{U, V_J\}$ has equivalent confounding strength to the observed covariates $\{X^{(b)}, K_J^{(b)} \}$ if
$$\frac{ \var(w(X, K_J, U, V_J) \mid A = 0)-\var(w(X, K_J) \mid A = 0)}{ \var(w(X, K_J, U, V_J) \mid A = 0)} =  \frac{ \var(w(X, K_J) \mid A = 0) - \var(w(\tilde X, \tilde K_J) \mid A = 0)}{\var(w(X, K_J) \mid A = 0)}.$$ Then, we can compute the corresponding $R^2$ parameter for a set of omitted confounders with equivalent confounding strength as $\{X^{(b)}, K_J^{(b)} \}$: 
$$R^2_{(b)} = \frac{\hat R^2_{-(b)}}{1+\hat R^2_{-(b)}}, \text{ where } \hat R^2_{-(b)} := 1-\frac{\var\left\{ w(\tilde X, \tilde K_J) \mid A = 0 \right\}}{\var \left \{ w(X, K_J) \mid A = 0 \right\} }$$
In settings where researchers restrict the set of benchmarked covariates to only individual-level covariates or only cluster-level covariates, then we can benchmark either the $R^2_U$ or $R^2_V$ value as appropriate. 

For the marginal sensitivity model, benchmarking requires assuming that omitting a confounder results in the same worst-case individual-level error as omitting an observed covariate. As such, researchers can find the corresponding $\Lambda$ parameter for an omitted confounder that results in the same maximum error as omitting $\{X^{(b)}, K_J^{(b)} \}$: 
$$\Lambda_{(b)} = \max \left\{\frac{w(\tilde X, \tilde K_J)}{w(X, K_J)}, \frac{w(X, K_J)}{w(\tilde X, \tilde K_J)} \right\}.$$

Benchmarking is most useful in settings when researchers have strong substantive priors for the types of covariates that are driving the underlying treatment assignment process. We caution that benchmarking cannot be used to rule out the existence of omitted confounders. However, it can be used to help reason about the plausibility of the magnitude of an underlying confounder. 

\subsection{Magnet Example}

Next, we include an amplification and benchmarking analysis for the Magnet Nursing application. We focus this analysis on the failure to rescue outcome for both the ATT and ATO estimands, since it is the largest effect in our analysis. This estimate was also the most robust to bias from a hidden confounder. We also restrict our analysis to the variance-based model, since use of these methods with the marginal model are nearly identical in form. We first implemented a benchmarking analysis. For this analysis, we estimated the weights while omitting some subset of the available covariates. Following the formula in Section~\ref{sec:bench}, we then estimate the benchmark for each subset of covariates. Given the nature of the imbalances at baseline, we focused on benchmarks for the cluster-level covariates.
For the patient-level covariates, we omitted the indicator for whether a patient was transferred as this covariate displayed the largest imbalance at baseline for any patient-level covariate. Finally, we estimated one set of weights where we omitted one cluster-level covariate and one patient-level covariate. For this set of weights, we omitted the indicator for a transfer patient and the percentage of cases that had surgery. 

\begin{figure}[htbp]
  \centering
    \includegraphics[scale=0.5]{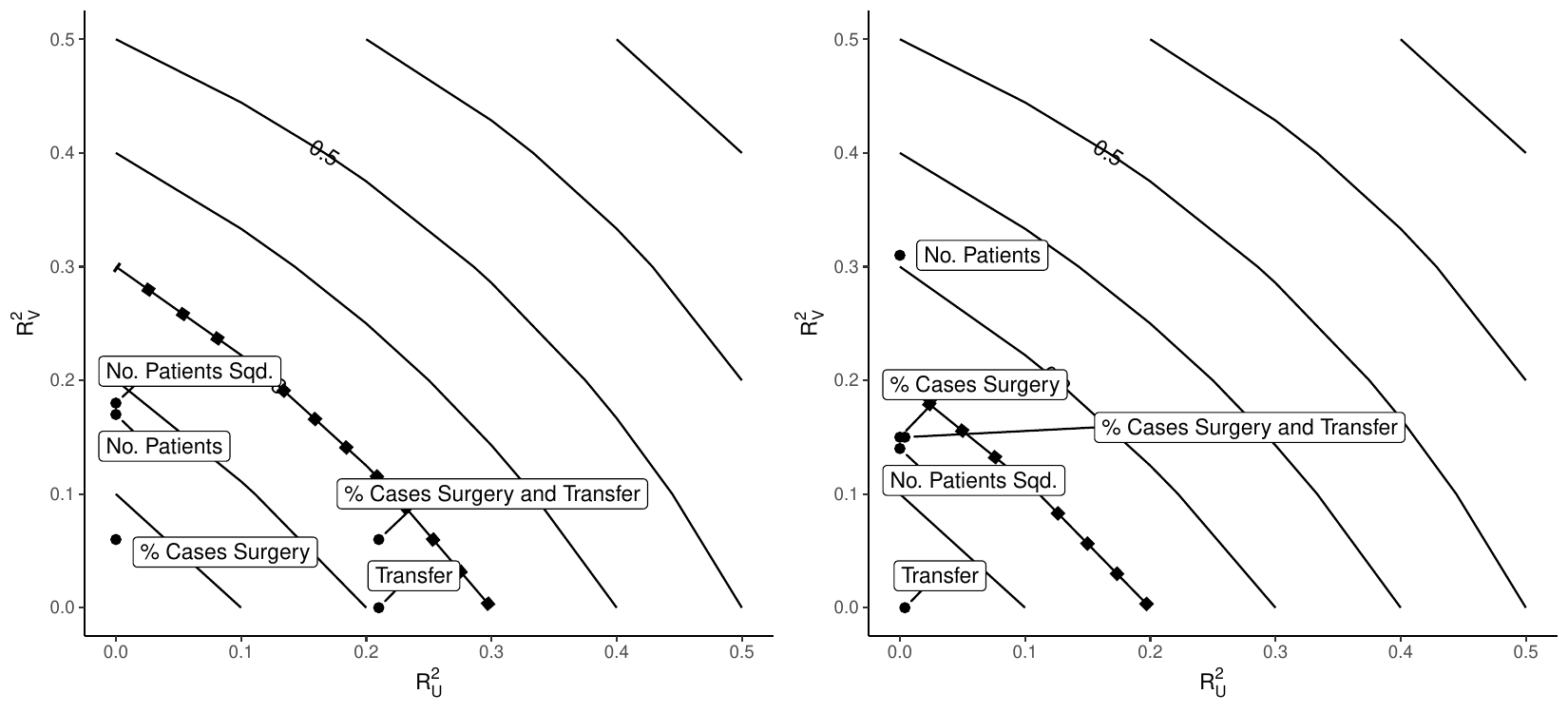}
    \caption{Benchmarking the sensitivity analysis for the variance-based sensitivity model. Dashed-line represents sensitivity analysis upper bound for failure to rescue outcome. The first panel contains results for the ATT estimand. The second panel contains the results for the ATO estimand.}
  \label{fig:amp}
\end{figure}

Figure~\ref{fig:amp} contains the results for both estimands. The demarcated line in the plots represents that values of $R^2_U$ and $R^2_V$
that result from the amplification of $R^2$ values of 0.4 and 0.3 calculated in Section~\ref{sec:mag}, representing the point where an unobserved confounder would alter our conclusions. The benchmark values in the plots represent the values where the observed covariates fall in terms of the $R^2_U$ and $R^2_V$ values. For the ATT estimand, none of the benchmark values exceed the $R^2$ threshold. This implies that a hidden confounder would have to have a larger effect than any of the observed confounders to change our conclusions. For the ATO estimand, only one covariate -- patient volume -- has a benchmark value that exceeds the $R^2$ value from the sensitivity analysis. This implies that the unobserved confounder would have to an effect on the weights that exceeds most of the observed confounders, but less than that of patient volume. Given that patient volume was an extremely strong predictor of Magnet status, this suggests that the unobserved confounder would have to be highly influential to change our substantive conclusions. In general, these results suggest that it would require a very strong unobserved confounder to change the study conclusions for the failure to rescue outcome.

\section{Conclusion}

In this paper, we developed sensitivity analysis methods for COS designs. Our framework contains several key developments. First, we developed methods for both marginal and variance based sensitivity models. Critically, in the COS setting the marginal model is far more conservative than the variance based methods, but does not yield sharp bounds. In addition, we extended our sensitivity analysis methods to the ATO estimand. The ATO estimand is commonly invoked in cases where overlap is poor and we have found that it is often necessary to deal with overlap violations in COS applications. To our knowledge, this is the first extension of sensitivity analysis methods for weighting estimators to the ATO. We also derived new bias decomposition results that are specific to the COS design. We show that imbalance in cluster-level covariates can amplify any bias in patient-level covariates, which suggests the importance of ensuring that adjustments for cluster-level covariates perform well.

This study also contributes to the literature on the effect of high-quality nursing on patient-level outcomes. Consistent with prior literature, we find that patients have better outcomes at Magnet Nursing hospitals but our study is the first to extend these findings to EGS conditions. Specifically, we found that patients had fewer adverse events and failures to rescue were also lower in Magnet Hospitals. Our study is also the first to use the COS framework to estimate the effect of Magnet Nursing on patient outcomes. We found that despite having a rich data set with a large number of hospitals, covariate overlap was limited, and we could only balance cluster-level covariates well once when we focused on the ATO estimand. Of course, this estimand limits the generalizability of the results. Focusing on the ATO implies that we are estimating the effect of Magnet Nursing certification for the set of hospitals that were on the margin of certification and have overlap between those with and without certification. Finally, we assessed whether these results were sensitive to bias from unobserved confounders. We found that the result for adverse outcomes could easily be explained by hidden confounding, while those for failure to rescue were fairly robust. Overall, the tools we develop here provide analysts with the means to implement sensitivity analyses for the COS design.

\clearpage
\renewcommand{\refname}{Bibliography}
\bibliography{cluster}

\clearpage
\noindent{\bfseries\Large Supplementary Materials}

\pagenumbering{arabic}
\setcounter{equation}{0}
\setcounter{table}{0}
\setcounter{figure}{0}
\renewcommand{\theequation}{S\arabic{equation}}
\renewcommand{\thetable}{S\arabic{table}}
\renewcommand{\thefigure}{S\arabic{figure}}

\appendix 

\section{Validity of Bias Decomposition for Overlap Weights}
Following \citet{huang2022sensitivity}, we must show that three conditions hold to apply the bias decomposition:
\begin{enumerate} 
\item $\E(w^* Y \mid Z = 0) = \mu$
\item $\E(w^*) = \E(w)$
\item $\E(w^* \mid X) = w$
\end{enumerate} 
Condition 1 implies that the ideal weights must recover our quantity of interest. Condition 2 means that the ideal and estimated weights must be centered at the same mean. This can be trivially met by normalization. Finally, Condition 3  is met when the projection of the ideal weights $w^*$ in the observed covariates $X$ recovers the estimated weights $w$. Condition 1 and 2 is met \citep{benmichael2022cos}. We thus show Condition 3. Let $w^* = P(Z_i = 1 \mid X, U)$, and $w* = P(Z_i = 1 \mid X)$.
\begin{align*} 
\E(w^* \mid X) &= \E \left[ P(Z_i = 1 \mid X, U) \mid X \right]\\
&= P(Z_i = 1 \mid X),
\end{align*} 
which follows directly by the tower property. 
\section{Bias decomposition for ATO} \label{app:bias_ato}
Let $\varepsilon_{V_J} = w(K_J, X) - w(K_J, X, V_J)$, and $\varepsilon_{U} =  w(K_J, X, V_J) - w(K_J, X, V_J, U)$. Furthermore, define $R^2_{V\mid A_{J} = a} := \var(\varepsilon_{V_J} \mid A_{J} = a)/\var(w(K_J, V_J, X) \mid A_{J} = a)$, and $R^2_{U \mid A_{J} = a} := \var(\varepsilon_{U} \mid A_{J} = a)/\var\left\{ w(K_J, V_J, X, U) \mid A_{J} = a\right\}$
Then, the bias decomposition for the ATO can be written as: 
\begin{align*} 
\text{Bias}(\hat \tau^{\textsc{ATO}}) 
= \sqrt{\frac{1}{1-R^2_{V|A_{J} = 1}}} & \Bigg\{ 
 \left(
\cor(\varepsilon_{V_J}, Y \mid A_{J} = 1) \sqrt{R^2_{V|A_{J} = 1}}+ \cor(\varepsilon_{U}, Y \mid A_{J} = 1)\cdot \sqrt{\frac{R^2_{U|A_{J} = 1}}{1-R^2_{U|A_{J} = 1}}} \right)\\
&~~~~~~~~~\times\sqrt{\var(Y\mid A_{J} = 1) \cdot \var\left\{ w(K_J, X)\mid A_{J} = 1\right\}} \Bigg\} \\
-\sqrt{\frac{1}{1-R^2_{V|A_{J} = 0}}} &\Bigg\{\left(
\cor(\varepsilon_{V_J}, Y \mid A_{J} = 0) \sqrt{R^2_{V|A_{J} = 0}}+ \cor(\varepsilon_{U}, Y \mid A_{J} = 0)\cdot \sqrt{\frac{R^2_{U|A_{J} = 0}}{1-R^2_{U|A_{J} = 0}}} \right)\\
&~~~~~~~~~~~~\times\sqrt{\var(Y\mid A_{J} = 0) \cdot \var\left\{ w(K_J, X)\mid A_{J} = 0\right\}} \Bigg\} 
\end{align*} 
\section{Proofs} 
\subsection{Proof of Theorem \ref{thm:bias}}
\begin{proof}
Let the weights estimated be $w(K_J, X)$. Let $w^*_i \equiv w(K_J, X, V_J, U)$. Now let an intermediary set of weights, in which we omit $U$, but not $V_J$ to be represented as $w(K_J, X, V_J)$. Then, the bias from omitting $\{V_J, U\}$ can be decomposed as follows. 
\begin{align*} 
\E[\hat \mu_W]&- \E[\hat \mu_W^*] \\
=& \cov(w(K_J, X) - w(K_J, X, V_J, U), Y) \\
=& \cov(w(K_J, X) - w(K_J, X, V_J) + w(K_J, X, V_J) - w(K_J, X, V_J, U), Y) \\
=&  \cov(w(K_J, X) - w(K_J, X, V_J), Y) + \cov(w(K_J, X, V_J) - w(K_J, X, V_J, U), Y) \\
=&\cor(\varepsilon_{V_J}, Y) \cdot \sqrt{\var(w(K_J, X) \cdot \frac{R^2_V}{1-R^2_V} \cdot \var(Y)} + 
\cor(\varepsilon_{U_{i}}, Y) \cdot \sqrt{\var(w(K_J, V_J, X)) \cdot \frac{R^2_U}{1-R^2_U} \cdot \var(Y)},
\intertext{where $R^2_V := \var(\varepsilon_{V_J})/\var(w(K_J, V_J, X))$, and $R^2_U := \var(\varepsilon_{U})/\var(w(K_J, V_J, X, U)$. We do not know what $\var(w(K_J, V_J, X))$ is. However, using our definition of $R^2_V$, we can re-write $\var(w(K_J, V_J, X)) = \var(w(K_J, X))/(1-R^2_V)$:}
=& \left( \cor(\varepsilon_{V_J}, Y) \cdot \sqrt{\var(w(K_J, X) \cdot \frac{R^2_V}{1-R^2_V}} + \cor(\varepsilon_{U_{i}}, Y) \cdot \sqrt{\var(w(K_J, V_J, X)) \cdot \frac{R^2_U}{1-R^2_U} }\right) \cdot \sqrt{\var(Y)}\\
=& \left( \cor(\varepsilon_{V_J}, Y) \cdot \sqrt{\var(w(K_J, X) \cdot \frac{R^2_V}{1-R^2_V}} + \cor(\varepsilon_{U_{i}}, Y) \cdot \sqrt{\frac{\var(w(K_J, X))}{1-R^2_V} \cdot \frac{R^2_U}{1-R^2_U} }\right) \cdot \sqrt{\var(Y)}\\
=& \sqrt{\frac{1}{1-R^2_V}} \cdot \left(\cor(\varepsilon_{V_J}, Y) \sqrt{R^2_V}+ \cor(\varepsilon_{U}, Y)\cdot \sqrt{\frac{R^2_U}{1-R^2_U}} \right) \cdot \sqrt{\var(Y) \cdot \var(w(K_J, X))}
\end{align*} 
\end{proof}
\subsection{Proof of Proposition \ref{prop:amplify_r2}}

\begin{align*} 
\frac{\var(w(X, K_J) \mid A = 0)}{\var(w^*(X, K_J, V_J, U) \mid A = 0)} &= 
\frac{\var(w(X, K_J) \mid A = 0)}{\var(w(X, K_J, V_J) \mid A = 0)} \cdot \frac{\var(w(X, K_J, V_J) \mid A = 0)}{\var(w(X, K_J, V_J, U \mid A = 0)}\\
&\leq (1-R^2_V) (1-R^2_U)\\
&= 1 - R^2_U - R^2_V + R^2_U \cdot R^2_V
\end{align*} 
Thus, when we use one $R^2$ value, this is equivalent to: 
$$R^2 = R^2_U \cdot R^2_V - (R^2_U + R^2_V)$$


\end{document}